\documentclass[conference]{IEEEtran}
\IEEEoverridecommandlockouts
\usepackage{cite}

\ifCLASSINFOpdf
\else
  \usepackage[dvips]{graphicx}
\fi
\usepackage[cmex10]{amsmath}
\usepackage{array}
\usepackage{url}
\usepackage{amsthm}
\usepackage{amssymb}
\usepackage{multirow}
\usepackage{dsfont}
\usepackage{graphicx}
\usepackage{color}
\theoremstyle{plain}

\theoremstyle{remark}

\begin{document}
\title{Coded Pilot Access: A Random Access Solution for Massive MIMO Systems}

\author{
\IEEEauthorblockN{
Jesper H. S\o rensen, Elisabeth de Carvalho, \v Cedomir Stefanovi\' c and Petar Popovski
}
\IEEEauthorblockA{
Aalborg University, Department of Electronic Systems,
Fredrik Bajers Vej 7, 9220 Aalborg, Denmark
\\
E-mail: \{jhs,edc,cs,petarp\}@es.aau.dk
\thanks{The research presented in this paper was partially supported by the Danish Council for Independent Research (Det Frie Forskningsr{\aa}d), grants no. DFF-1335-00273 and DFF-4005-00281. Part of this work has been performed in the framework of the Horizon 2020 project FANTASTIC-5G (ICT-671660), which is partly funded by the European Union. The authors would like to acknowledge the contributions of their colleagues in FANTASTIC-5G.}
}
}
\maketitle

\begin{abstract}
We present a novel access protocol for crowd scenarios in massive MIMO (Multiple-input multiple-output) systems. Crowd scenarios are characterized by a large number of users with intermittent access behavior, whereby orthogonal scheduling is infeasible. In such scenarios, random access is a natural choice. The proposed access protocol relies on two essential properties of a massive MIMO system, namely asymptotic orthogonality between user channels and asymptotic invariance of channel powers. Signal processing techniques that take advantage of these properties allow us to view a set of contaminated pilot signals as a graph code on which iterative belief propagation can be performed. This makes it possible to decontaminate pilot signals and increase the throughput of the system. Numerical evaluations show that the proposed access protocol increases the throughput with $36\%$, when having $400$ antennas at the base station, compared to the conventional method of slotted ALOHA. With $1024$ antennas, the throughput is increased by $85\%$.
\end{abstract}
\IEEEpeerreviewmaketitle

\section{Introduction} \label{sec:introduction}
Massive MIMO (Multiple-input multiple-output) has been identified as a key technology to improve spectral efficiency of wireless communication systems and one of the main enablers of the upcoming 5th generation \cite{Petar5G}. A massive MIMO system refers to a multi-cell multi-user system with a massive number of antennas at the BS that serves multiple users~\cite{marzetta06}. The number of users is much smaller than the number of BS antennas, defining an under-determined multi-user system with a massive number of extra spatial degrees of freedom (DoF). Exploiting those extra DoF and assuming an infinite number of antennas at the BS, the multi-user MIMO channel can be turned into an orthogonal channel and the effects of small-scale fading and thermal noise can be eliminated. 

However, when the number of antennas becomes massive, acquiring the channel state information (CSI) becomes a severe bottleneck. Downlink channel training requires a training length that is proportional to the number of antennas at the BS and is thus impractical. A solution promoted in \cite{marzetta06} restricts massive MIMO operations to time-division duplex (TDD) for which channel reciprocity is exploited. As the downlink and uplink channels are equal, CSI is acquired at the BS based on uplink training and then used for downlink transmission. The benefit is that the training length is proportional to the number of users, which is much smaller than the number of BS antennas. 

As described in \cite{marzetta06}, CSI is acquired using orthogonal pilot sequences, but, due to the shortage of orthogonal sequences, the same pilot sequences must be reused in neighboring cells, causing pilot contamination. This problem is considered as one of the major challenges in massive MIMO systems \cite{rusek13}. Mitigation of pilot contamination has been the focus of several works recently. These include \cite{gesbert13}, where it is utilized that the desired and interfering signals can be distinguished in the channel covariance matrices, as long as the angle-of-arrival spreads of desired and interfering signals do not overlap. A pilot sequence coordination scheme is proposed to help satisfying this condition. The work in \cite{ashikhmin12} utilizes coordination among base stations to share downlink messages. Each BS then performs linear combinations of messages intended for users applying the same pilot sequence. This is shown to eliminate interference when the number of base station antennas goes to infinity. A multi-cell precoding technique is used in \cite{jose11} with the objective of not only minimizing the mean squared error of the signals within the cell, but also minimizing the interference imposed to other cells. 

The pilot contamination problem has been seen as an inter-cell problem that arises when the users associated with two neighboring cells use the same pilot sequence. An implicit assumption associated with it is that the pilot sequences of the users associated with the same cell are perfectly scheduled, such that no intra-cell pilot contamination occurs. These assumptions fall apart when one considers very dense, crowd scenarios as envisioned in 5G wireless scenarios \cite{METIS}. In such a setting, orthogonal scheduling of the users belonging to the same BS becomes infeasible due to scheduling overhead. 

This work is motivated by the massive MIMO problem in a crowd setting, as well as the observation that the pilot contamination problem is very much dependent on the \emph{protocol assumptions} made in the system. 
Specifically, we consider a crowd scenario where the amount of users and their access behavior make it infeasible to schedule the transmissions. Instead users choose pilot sequences at random in an uncoordinated manner from a small pool shared by all users. In this way, the inter-cell pilot contamination problem becomes an intra-cell pilot contamination problem, where the BS needs to handle collisions that occur in the pilot domain. Some recent works have considered this approach. The paper \cite{Icassp} features a proposal for a joint pilot and data transmission in the uplink, where the purpose is a reliable communication by following an ergodic process. In each time slot, a user selects uniformly at random a pilot sequence and selects part of her codeword. The packet collisions are neither detected nor resolved, while over an asymptotically long time horizon, fading and effects of pilot contamination are averaged out allowing the determination of a reliable rate for transmission. Another related work is \cite{SUCR}, where the pilots are transmitted by using the random access procedure in LTE, but modified according to the specifics of pilot access. The paper proposes an approach to resolve one-shot collisions by exploiting the channel hardening properties of massive MIMO and enabling the terminals to detect the collision and act accordingly. 

The approach devised in this paper differs from \cite{Icassp} and \cite{SUCR} by having the terminals participate in a \emph{coded random access} procedure \cite{Liva,SPV2012}. The terminals transmit with a predefined probability and send a pilot in the uplink followed by the data part. Pilot assignment is randomized in each time slot while the data part is repeated, which enables to use successive interference cancellation across the replicas of the same packet. One of the main contributions is the way successive interference is implemented, as it relies on two features specific to massive MIMO: (1) asymptotic orthogonality between user channels; and (2) asymptotic invariance of the power received from a user over a short time interval. Furthermore, we also propose to have flexible switching from uplink to downlink transmission on a time slot basis for the users that are not affected by pilot collision. 

A preliminary version of this work has appeared in \cite{gcws}. The present version provides a thorough elaboration on the and-or tree analysis of the degree distribution of the random access code. Differently from \cite{gcws}, this work considers a channel code at the physical layer, which exists in most practical systems, and is shown to greatly influence the design and performance of the random access code.




\section{System Model}\label{sec:sysmodel}
In this work we denote scalars in lower case, vectors in bold lower case and matrices in bold upper case. A superscript `$T$' denotes the transpose, a superscript `$*$' denotes the complex conjugate and a superscript `$H$' denotes the conjugate transpose.

\begin{figure}[t]
 \centering
 \includegraphics[width=1\columnwidth]{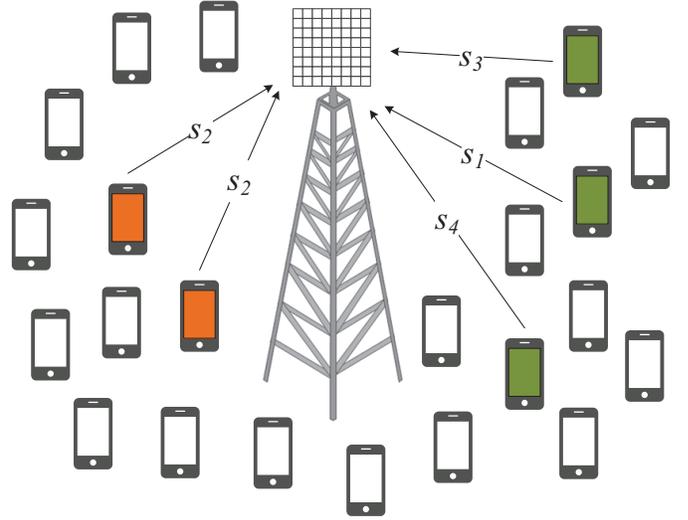}
 \caption{A single cell crowd scenario. Red devices experience interference due to colliding pilot signals. Green devices apply unique pilot sequences, whereby interference is avoided.}
 \label{fig:system}
\end{figure}

\begin{figure}[t]
 \centering
 \includegraphics[width=1\columnwidth]{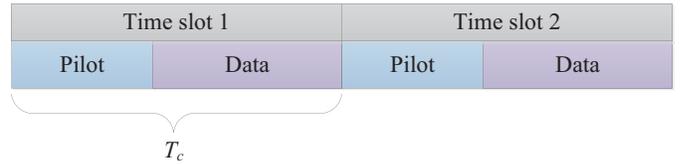}
 \caption{An example of a transmission schedule.}
 \label{fig:transschedule}
\end{figure}

\begin{figure}[t]
 \centering
 \includegraphics[width=1\columnwidth]{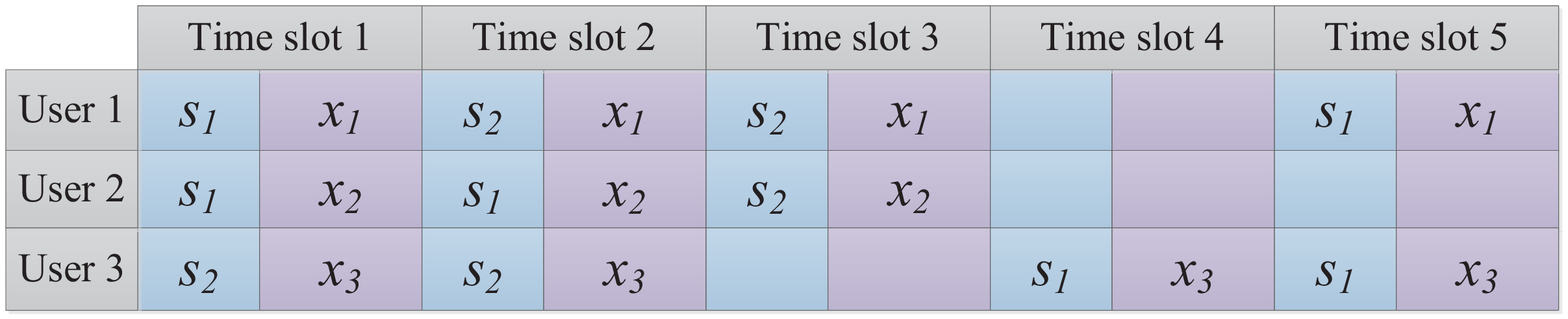}
 \caption{An example of a pilot schedule. Subscripts for pilots refer to indexing within the set of pilots, whereas subscripts for data refer to users.}
 \label{fig:pilotschedule}
\end{figure}

We consider a random access system consisting of a single base station with $M$ antennas and $K$ users, each one with a single antenna, see Fig.~\ref{fig:system}. Communication is performed by using slotted time, where each time slot consists of an uplink pilot phase and a data phase (either uplink or downlink), see Fig.~\ref{fig:transschedule}. In each time slot, each user is active with probability $p_a$. 
There are $\tau$ orthogonal pilot sequences $\{\pmb{s}\}$, each consisting of $\tau$ symbols $\pmb{s}=\left[s(1)\hspace{0.1cm}s(2) \hdots s(\tau)\right]$. 
An active user selects a pilot sequence randomly from the $\tau$ available pilot sequences. Note that multiple users may choose the same pilot sequence. See Fig.~\ref{fig:pilotschedule} for an example of a random pilot schedule with $\tau=2$ and $K=3$.
The channel between the $k$'th user and the BS during the $n$'th time slot is denoted $\pmb{h}_{nk}=\left[h_{nk}(1)\hspace{0.1cm}h_{nk}(2) \hdots h_{nk}(M)\right]^T$, where $h_{nk}(i) \sim \mathcal{CN}(0,1)$, $\forall \, i$. The time slot has a duration $T_c$, which corresponds to the coherence time in which the channel coefficient remains constant. Channel coefficients in different time slots are assumed i.i.d. 
Let $\mathcal{A}_n$ denote all active users in time slot $n$, while $\mathcal{A}_n^j$ denotes the set of users that have selected $\pmb{s}_j$ in the $n$'th time slot. If $\pmb{Y}_{n}^{pu} \in {\cal C} ^{M \times \tau}$, denotes the uplink pilot signal received in time slot $n$, we have
\begin{align}\label{eq:pilots}
\pmb{Y}_{n}^{pu} = \sum_{j=1}^{\tau} \sum_{k\in\mathcal{A}_n^j} \pmb{h}_{nk} \pmb{s}_j + \pmb{Z}_{nj}^{pu},
\end{align}
\noindent where $\pmb{Z}_{nj}^{pu}$ is a matrix of i.i.d. Gaussian noise components, hence $\pmb{Z}_{nj}^{pu}(i,j) \sim \mathcal{CN}(0,\sigma_n^2)$, $\forall \, i,j$. Any future instances of a vector $\pmb{z}$ or matrix $\pmb{Z}$, with different sub- or superscripts follow the same definition. All active users transmit a message of length $D$ symbols in the uplink data phase. The message from the $k$'th user is denoted $\pmb{x}_k^u=\left[x_k^u(1)\hspace{0.1cm}x_k^u(2) \hdots x_k^u(D)\right]$. Using $\pmb{Y}_{n}^u \in {\cal C} ^{M \times D}$, to denote the data part of the received signal in the uplink, we get:
\begin{align}\label{eq:udata}
\pmb{Y}_{n}^u = \sum_{k\in\mathcal{A}_n} \pmb{h}_{nk} {\pmb{x}_k^u} + \pmb{Z}_n^u.
\end{align}
In the downlink phase we rely on channel reciprocity, such that the uplink channel estimate is assumed to be a valid estimate for the downlink transmission. The BS transmits a precoded downlink pilot sequence, such that the $k$-th user receives a downlink pilot signal, $\pmb{y}_{nk}^{pd} \in {\cal C} ^{1 \times \tau}$, given by
\begin{align}\label{eq:pilots_down}
\pmb{y}_{nk}^{pd} = \pmb{h}_{nk}^T \pmb{w}_{nk} \pmb{s}_j + \pmb{z}_{nk}^{pd},
\end{align}
\noindent where $\pmb{w}_{nk}=\left[w_{nk}(1)\hspace{0.1cm}w_{nk}(2) \hdots w_{nk}(M)\right]^T = \pmb{h}_{nk}^*$ is the precoding vector for user $k$ in the $n$'th time slot. Clearly, this assumes that the BS has an estimate of the channel $\pmb{h}_{nk}$ before the downlink transmission. The BS is able to schedule the downlink messages, $\pmb{x}_k^d=\left[x_k^d(1)\hspace{0.1cm}x_k^d(2) \hdots x_k^d(D)\right]$, such that the received downlink data signal is
\begin{align}\label{eq:ddata}
\pmb{y}_{nk}^d = \pmb{h}_{nk}^T \pmb{w}_{nk} {\pmb{x}_k^d} + \pmb{z}_{nk}^{d}.
\end{align}
In both uplink and downlink, the coherence time allows the transmission of $L$ symbols and we have $L=\tau + D$. Furthermore, the data is assumed to be channel coded with rate $R$ at the physical layer, such that the effective data rate is $R\frac{D}{L}$. By $\mathcal{S}_n$ we denote the set of users, whose associated data message, uplink or downlink, is successfully recovered. Note that $|\mathcal{S}_n| \le \tau$ and $|\mathcal{S}_n| \le |\mathcal{A}_n|$. The uplink/downlink throughput of the system in time slot $n$, $\gamma_n$, is then defined as the sum-rate given by
\begin{align}\label{gamman}
\gamma_n = \frac{|\mathcal{S}_n| R (L-\tau)}{L}=\frac{|\mathcal{S}_n| R D}{L}
\end{align}
%


\section{Coded Pilot Access}\label{sec:scheme}
This section describes the proposed method of communication in the system described in Section \ref{sec:sysmodel}, treating both uplink and downlink operation.

\subsection{Uplink}
\label{sec:scheme_uplink}
In uplink operation, transmissions are organized in blocks of $\Delta$ consecutive time slots, referred to as a \textit{frame}. If a user is active multiple times within a frame, the uplink data is retransmitted, similar to conventional random access schemes. We introduce a parameter called the \textit{overhead factor} $\alpha$, defined as 
\begin{align}
\label{eq:alpha}
\alpha= \frac{\tau \Delta}{K},
\end{align}
which is an expression of the normalized amount of orthogonal resources in a frame. The performance parameter of interest is the frame average throughput given by $\gamma=\sum_{n=1}^\Delta \gamma_n / \Delta$. From the uplink pilot signals in \eqref{eq:pilots}, it is possible to estimate the channels between the users and the base station. However, since multiple users may apply the same pilot sequence, it is only possible to estimate a sum of the involved channels. The least squares estimate, $\pmb{\phi}_{nj}$, based on the pilot signal in time slot $n$ from users applying $\pmb{s}_j$ is found as
\begin{align}\label{eq:estimates}
\pmb{\phi}_{nj} &= (\pmb{s}_j \pmb{s}_j^H)^{-1} {\pmb{Y}_{n}^{pu}} \pmb{s}_j^H \notag \\
       &= \sum_{k\in\mathcal{A}_n^j} \pmb{h}_{nk} + \pmb{z}_{nj}^{pu'}.
\end{align}
\noindent where $\pmb{z}_{nj}^{pu'}$ is the post-processed noise terms originating from $\pmb{z}_{nj}^{pu}$. Any future instances of a vector $\pmb{z}$ with a prime follow the same definition.

The problem of interfering users applying the same, or a non-orthogonal, pilot sequence is often called \textit{pilot contamination}. If we proceed to detect the data in the uplink phase using a contaminated channel estimate, the result will be a summation of data messages. If $\pmb{\psi}_{nj}$ is the data estimate based on the channel estimate $\pmb{\phi}_{nj}$, we have
\begin{align}
\pmb{\psi}_{nj} &= (\pmb{\phi}_{nj}^H \pmb{\phi}_{nj})^{-1} \pmb{\phi}_{nj}^H \pmb{Y}_{n}^u \notag \\
&= \sum_{k\in\mathcal{A}_n^j} \frac{\pmb{\phi}_{nj}^H\pmb{h}_{nk}}{||\pmb{\phi}_{nj}||^2} \pmb{x}_{k}^u + \pmb{z}_{n}^{u'}.
\end{align}
Hence, a pilot collision leads to a data collision. A classical way to deal with this problem is to minimize the probability of contamination by carefully selecting $p_a$. The objective of such criterion is to maximize the probability of having only one user applying a particular pilot sequence in a particular time slot. Hence, we have
\begin{align}\label{eq:max}
\underset{p_a}{\text{maximize}} & & \text{Pr}(\left\vert{\mathcal{A}_n^j}\right\vert = 1) \notag \\
\text{subject to} & & 0 \le p_a \le 1
\end{align}
This will maximize the number of non-contaminated channel estimates, and in turn maximize the number of successful data transmissions. This approach is reminiscent of the framed slotted ALOHA protocol for conventional random access. We consider this a reference scheme in this work and refer to it as ALOHA. Note that a random access, i.e. nonscheduled scheme must be considered as a reference, due to the assumption of a crowd scenario, where scheduling is infeasible.

A novel alternative solution is presented in this paper, which does not consider data collisions as waste, but instead buffers the collided signals and use them subsequently through an iterative process, whereby they contribute to the throughput. We call it \textit{Coded Pilot Access} (CPA). This solution is based on applying the contaminated estimates as \textit{matched filters} on the received uplink data signals, $\pmb{Y}_{n}^u$. Denoting the filtered data signal $\pmb{f}_{nj}\in {\cal C} ^{1 \times D}$, we have
\begin{align}\label{eq:fnj1}
\pmb{f}_{nj} &= \pmb{\phi}_{nj}^H \pmb{Y}_{n}^u \notag \\
&=\sum_{k\in\mathcal{A}_n^j} \left( ||\pmb{h}_{nk}||^2 + \sum_{m\in\mathcal{A}_n^j\backslash \{k\}} \pmb{h}_{nm}^H \pmb{h}_{nk} \right) \pmb{x}_{k}^u \notag \\
&\quad + \sum_{\ell\in \mathcal{A}_n \backslash \mathcal{A}_n^j} \left( \sum_{o\in\mathcal{A}_n^j} \pmb{h}_{no}^H \pmb{h}_{n\ell} \right) \pmb{x}_{\ell}^u + \pmb{z}_{n}^{u'}.
\end{align}
By relying on two essential features from the massive MIMO scenario, \eqref{eq:fnj1} can be simplified greatly, when $M$ goes towards infinity. The first feature is orthogonality between user channel vectors. This implies that $\pmb{h}_{nm}^H \pmb{h}_{nk}=0$ for $M \to \infty$. The second feature is the temporal stability of channel powers, which implies that $||\pmb{h}_{nk}||^2=||\pmb{h}_{n'k}||^2$ $\forall$ $n,n'$. This allows us to drop the time index in the channel powers. We thus have the following expression for the filtered data signal in the limit of $M \to \infty$:
\begin{align}\label{eq:fnj2}
\lim_{M \to \infty} \pmb{f}_{nj} &=\sum_{k\in\mathcal{A}_n^j} ||\pmb{h}_{k}||^2 \pmb{x}_{k}^u + \pmb{z}_{n}^{u'}.
\end{align}
Hence, the implications of pilot contamination has been turned into linear combinations of data messages, through post-processing with matched filters. The coefficients of the linear combinations are the temporally stable channel powers. By again relying on the asymptotic properties of the massive MIMO channel, the channel vector estimates in \eqref{eq:estimates} can be utilized to find estimates of the sums of the channel powers. We denote these as $\pmb{g}_{nj}$ and have
\begin{align}\label{eq:gnj}
\pmb{g}_{nj} &= \pmb{\phi}_{nj}^H \pmb{\phi}_{nj} \notag \\
&=\sum_{k\in\mathcal{A}_n^j} \left( ||\pmb{h}_{nk}||^2 + \sum_{m\in\mathcal{A}_n^j\backslash \{k\}} \pmb{h}_{nm}^H \pmb{h}_{nk} \right) + \pmb{z}_{nj}^{pu'}, \notag \\
\lim_{M \to \infty} \pmb{g}_{nj} &=\sum_{k\in\mathcal{A}_n^j} ||\pmb{h}_{k}||^2 + \pmb{z}_{nj}^{pu'}.
\end{align}
Eqs. \eqref{eq:fnj2} and \eqref{eq:gnj} for $n=1,\hdots, \Delta$, and $j=1,\hdots,\tau$, represent a system of equations, which we wish to solve for $\pmb{x}_{k}^u$, $k=1,\hdots,K$. It should be noted that the BS has no a-priori knowledge of the random activity and pilot choices of the users. Hence, the system of equations cannot be solved using, e.g. Gaussian elimination. Instead we employ \textit{successive interference cancellation} (SIC), as in recent works on CRA \cite{PSLP2014}. 

SIC proceeds as follows. Initially, the BS locates immediately decodable uplink data\footnote{In practice this is enabled by applying a cyclic redundancy check (CRC) code on the data.}, i.e. cases of $\left\vert{\mathcal{A}_n^j}\right\vert = 1$. If $\pmb{f}_{nj}$ is decodable, we furthermore have an estimate of the channel norm of the transmitter in $\pmb{g}_{nj}$. Embedded in the uplink data is the random activity and pilot choices of the transmitter\footnote{A practical solution to this is to embed the seed for the random number generator. The rate loss due to the CRC and the embedded seed is considered negligible.}, which allows the BS to locate all the replicas of the same packet sent by that transmitter. In the context of \eqref{eq:fnj2} and \eqref{eq:gnj}, when the data from user $k$ is successfully decoded, the BS learns for which $n$ and $j$ we have $k\in\mathcal{A}_n^j$. This enables the BS to cancel the interference caused by the replicas from user $k$ by subtracting $||\pmb{h}_{k}||^2 \pmb{x}_{k}^u$ from any $\pmb{f}_{nj}$ for which $k\in\mathcal{A}_n^j$. Furthermore, the interference caused by the associated pilot transmissions can be canceled by subtracting $||\pmb{h}_{nk}||^2$ from any $\pmb{g}_{nj}$ for which $k\in\mathcal{A}_n^j$. The cancellations cause $k$ to be removed from any $\mathcal{A}_n^j$ it originally appeared in. Potentially, this leads to new cases of $\left\vert{\mathcal{A}_n^j}\right\vert = 1$, whereby new data can be recovered, and the iterative process can continue. 

\begin{figure}[t]
 \centering
 \includegraphics[width=1\columnwidth]{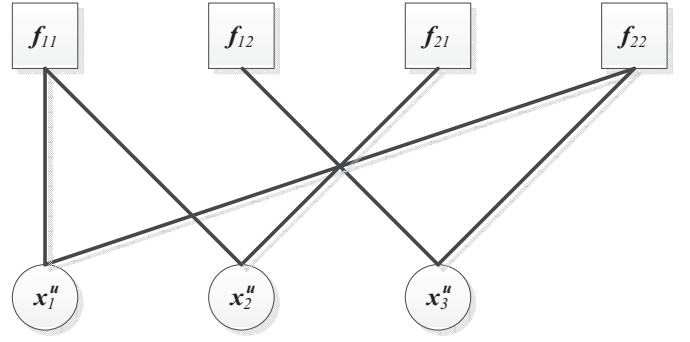}
 \caption{A bipartite graph representation of the data collisions.}
 \label{fig:graphf}
\end{figure}

The employed decoding algorithm is analogous to belief propagation (BP) decoding of erasure codes. A common way of visualizing such codes is by using bipartite graphs. They also apply in our context, see Fig.~\ref{fig:graphf} for an example based on the first two time slots in the example from Fig.~\ref{fig:pilotschedule}. Squares are referred to as factor nodes and represent observable signals after matched filtering. Hence, each factor node corresponds to an orthogonal resource, i.e. a pilot in a time slot within a frame. Circles are referred to as variable nodes and represent data messages, which we wish to recover. An edge connecting a variable node with a factor node means that the variable is a part of the linear combination represented by the factor node. The number of edges connected to a node is referred to as the degree of the node. Based on Fig.~\ref{fig:graphf}, we can walk through the simple example of decoding $\pmb{x}_{1}^u$.

\textbf{Example:} Initially the BS detects that $\pmb{f}_{12}$ has degree one, and thereby directly recovers $\pmb{x}_{3}^u$. The data from user $3$ makes the BS aware of the activity pattern of this user and thereby enables cancellation of its interference. As a result, $||\pmb{h}_{3}||^2$ is subtracted from $\pmb{g}_{22}$ and $||\pmb{h}_{3}||^2 \pmb{x}_{3}^u$ is subtracted from $\pmb{f}_{22}$. This cancellation has reduced the degree of $\pmb{f}_{22}$ to one, which is detected by the BS, whereby $\pmb{x}_{1}^u$ is recovered.

In the example, noise is assumed to not garble the decoding of the rate $R$ channel code applied at the physical layer. In other words, when a signal has been reduced to degree one, the corresponding data is recovered successfully. Clearly, this is a strong assumption that cannot always hold. In fact, signals of higher degree have a higher risk of being undecodable at the physical layer after being reduced to degree one. The reason is accumulation of noise during interference cancellation. While interference cancellation greatly increases the SINR, it actually decreases the SNR, which potentially makes a rate $R$ channel code undecodable. This effect is analyzed in section \ref{sec:analysis}. 

The performance of the BP decoder for erasure codes is tightly connected with the factor- and variable-node \textit{degree distribution}, denoted as $\Psi$ and $\Lambda$, respectively, where $\Psi_d$/$\Lambda_d$ is the probability that a factor/variable node has degree $d$. Several works \cite{lt,sorensen} have studied the design of well performing degree distributions. However, in this context we do not have the full freedom to tailor the degree distributions. Our only way of influencing degree distributions is through the choice of $p_a$ and the overhead factor $\alpha$, see \eqref{eq:alpha}. Specifically, since a user is applying a particular pilot sequence in a particular time slot with probability $p_a/\tau$, and there are $\Delta$ time slots, we have the following relation between the degree distributions and $p_a$ and $\alpha$:
\begin{align}
\Psi_d&=\text{Pr}(\left\vert{\mathcal{A}_n^j}\right\vert = d) = \binom{K}{d}\left(\frac{p_a}{\tau}\right)^d\left(1-\frac{p_a}{\tau}\right)^{K-d} \\
\label{eq:degree}
& \approx \frac{( \frac{p_a K}{\tau} )^d}{d!} e^{- \frac{p_a K}{\tau} }  = \frac{ \beta^d}{d!} e^{- \beta},
\end{align}
where $\beta$ is the average factor-node degree
\begin{align}\label{eq:bard}
\beta = \frac{ p_a K}{\tau}
\end{align}
and
\begin{align}
\label{eq:big_lambda}
\Lambda_d & = { \Delta \choose d } p_a^d ( 1 - p_a )^{\Delta - d} \approx \frac{ (\Delta \, p_a )^d }{d!} e^{ - \Delta p_a } \\
& = \frac{ (\alpha \beta ) ^d}{d!} e^{- \alpha \beta}.
\end{align}
Obviously, through the choice of $\beta$ (i.e. the choice of $p_a$) and $\alpha$, one determines the degree distributions. In section \ref{sec:analysis} we provide the analytical optimization of $\beta$ and $\alpha$.

\subsection{Downlink}
In order to choose an appropriate precoder for the downlink transmission, the BS must have an estimate of the current channel. The coded operation applied in uplink, which results in multiple collisions and occasional single (non-contaminated) transmissions, does not guarantee that such an estimate is available. Uplink operation relies on SIC based only on knowledge of the norm. Hence, downlink transmission to a user is only possible if that user avoided collision during the previous uplink pilot phase, such that an uncontaminated channel estimate is available. This incurs a delay in downlink transmissions, which we denote as $\Delta_k$ for user $k$. This delay is equal to the number of time slots until user $k$ is active and avoids a collision during the uplink pilot phase. Denoting the probability of a user being active and avoiding collision, $p_a'$, we have 
\begin{align}\label{eq:delay}
p_a' = p_a\left(1-\frac{p_a}{\tau}\right)^{K-1}.
\end{align}
\noindent The probability distribution of $\Delta_k$ is the geometric distribution and is therefore given by
\begin{align}\label{eq:delay_dist}
Pr(\Delta_k=\delta) = p_a' (1-p_a')^{\delta-1}.
\end{align}
\noindent The expected value, $\mathrm{E}[\Delta_k]$, of the delay is then found as
\begin{align}\label{eq:delay_exp}
\mathrm{E}[\Delta_k] = \frac{(1-p_a')}{p_a'}.
\end{align}
\noindent There is a natural tradeoff between optimizing $p_a$ for high uplink throughput and optimizing it for limiting the delay in the downlink phase. Such a joint optimization is outside the scope of this work. In the numerical evaluations in section \ref{sec:results}, we will solely be concerned with the uplink throughput.

Regarding the reception of a downlink transmission and assuming channel reciprocity, the $k$'th user does not need to estimate each coefficient of $\pmb{h}_{nk}$, which would require a pilot signal for all $M$ antennas. Instead, we let the receiver estimate the concatenated ``channel'' consisting of both the downlink precoder, $\pmb{w}_{nk}$, and the actual channel. Denoting the concatenated channel, $q_{nk}$, we have
\begin{align}\label{eq:down}
q_{nk} &= \pmb{h}_{nk}^T \pmb{w}_{nk},
\end{align}
\noindent where $q_{nk}$ is estimated through \eqref{eq:pilots_down}.


\section{Analysis} \label{sec:analysis}

The SIC algorithm described in Section~\ref{sec:scheme} can be analyzed using the analytical tools devised for BP erasure decoding, specifically, using the and-or tree evaluation \cite{LMS1998}.\footnote{For a general introduction to the and-or tree evaluation, we refer the interested reader to \cite{LMS1998,MCT}.}
For the given factor node degree distributions $\Psi$ and variable node degree distribution $\Lambda$, the and-or tree evaluation outputs the asymptotic probability, when $K \rightarrow \infty$, of recovering a user signal. 
However, there are important differences to the standard and-or tree evaluation that have to be taken into account, stemming from the nature of the physical layer operation:
\begin{itemize}
\item The decodability of user signals received in singleton slots depends on the received SNR.
\item The cancellation of decoded signals is not ideal and leaves residual interference power. This implies that, as the SIC progresses, the accumulated residual interference effectively decreases SNR, which may prevent decoding of the signals whose degree become reduced to one.
\end{itemize}

\begin{figure}[t]
 \centering
 \includegraphics[width=0.8\columnwidth]{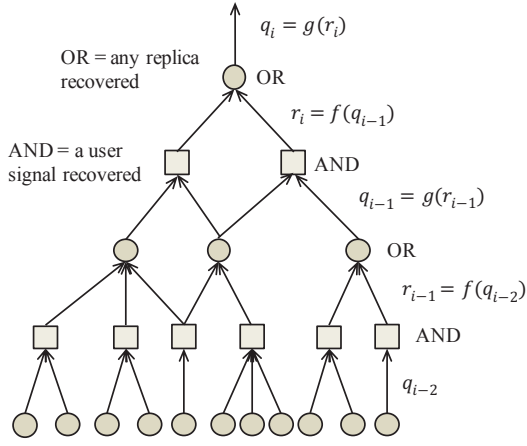}
 \caption{A tree representation of the iterative recovery of user signals.}
 \label{fig:tree}
\end{figure}

The and-or tree evaluation assumes that the bipartite graph representation, Fig.~\ref{fig:graphf}, can be unfolded into a tree, see Fig.~\ref{fig:tree}, on which two operations are iteratively performed:
\begin{itemize}
\item[(i)] recovery of user signals in factor nodes, corresponding to (a generalized) ``and'' operation, cf. \cite{Liva,SMP2014},
\item[(ii)] removal of replicas of decoded signals, corresponding to the ``or'' operation.
\end{itemize}
Both operations are probabilistically characterized, in terms of probability of \emph{not} decoding a user in a slot, denoted as $r_{i}$, and \emph{not}
removing a replica, denoted as $q_{i}$, respectively.
The tree structure allows for their successive updates, as depicted in Fig.~\ref{fig:tree}.
We note that in the non-asymptotic case, the graph representation contains loops, and the corresponding tree representation is only an approximation, where the obtained results present an upper bound on the non-asymptotic performance.

Before providing the expressions for $r_{i}$ and $q_{i}$, we introduce the edge-oriented degree distributions \cite{LMS1998}, corresponding to probabilities that a randomly chosen edge in the graph is connected to a node of a certain degree.
In particular, there are the edge-oriented factor-node degree distribution $\psi$ and the edge-oriented variable-node degree distribution $\lambda$, which can be derived through the factor- and variable-node degree distributions $\Psi$ and $\Lambda$, respectively \cite{LMS1998}
\begin{align}
\label{eq:psi}
\psi_d = \frac{d \, \Psi_d}{\sum_j j \, \Psi_j}, \; d \geq 1, \\
\label{eq:lambda}
\lambda_d = \frac{d \, \Lambda_d}{\sum_j j \, \Lambda_j}, \; d \geq 1.  
\end{align}

\begin{figure}[t]
 \centering
 \includegraphics[width=0.7\columnwidth]{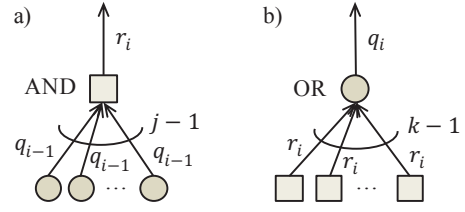}
 \caption{Probability updates in a) factor node and b) variable node.}
 \label{fig:ops}
\end{figure}

Assume a factor node of degree $j$.
The probability that an edge connected to a factor node of degree $j$ is not removed in the $i$'th iteration, denoted by $r_{i|j}$, is
\begin{align}
r_{i|j} = \pi_{j} \, (1-q_{i-1})^{j-1}, \; i \geq 1,  %
\end{align}
where $\pi_{j}$ is the probability of recovering a user signal in the factor node of degree $j$ when $j-1$ interfering signals
were cancelled\footnote{I.e. the probability of recovering a user signal from a factor node whose original degree $j$ is reduced to 1.}, and 
the term $(1-q_{i-1})^{j-1} $ refers to the probability that $j-1$ interfering signals were cancelled, see Fig.~\ref{fig:ops}\,a).
It is important to note that the impact of the physical layer, i.e. receiver operation, as described in Section~\ref{sec:scheme}, is embedded in $\pi_{j}$.\footnote{In the standard and-or tree evaluation, $\pi_j =1 $, $j \geq 1$.}
Averaging over the edge-oriented factor-node degree distribution yields
\begin{align}
\label{eq:r_i}
r_{i} = \sum_j  \psi_{j} \, r_{i|j}  = \sum_j \psi_{j} \, \pi_{j} \, (1-q_{i-1})^{j-1},\; i\geq1.
\end{align}
Further, the probability that an edge connected to a variable node of degree $k$ is not removed in the $i$'th iteration, denoted by $q_{i|k}$, is
\begin{align}
q_{i|k}=\sum_{k} r_{i}^{k-1},\;i\geq1,
\end{align}
where $r_{i}^{k-1}$ refers to the probability that none of the $k-1$ replicas were recovered, see Fig.~\ref{fig:ops}\,b). 
Averaging over the edge-oriented variable-node degree distribution produces
\begin{align}
\label{eq:q_i}
q_{i}= \sum_{k} \lambda_{k} \, q_{i|k} = \sum_{k} \lambda_{k} \, r_{i}^{k-1},\;i\geq1,
\end{align}
with the initial value $q_{0}=1$.
Combining \eqref{eq:r_i} and \eqref{eq:q_i}, we get
\begin{align}
\label{eq:composed}
q_i = \sum_{k} \lambda_{k} \Big( \sum_j \psi_{j} \, \pi_{j} \, (1-q_{i-1})^{j-1} \Big), \; i \geq 1.
\end{align}
The output of the evaluation is the probability that a user signal becomes recovered:
\begin{align}
p_{d} = 1 - \lim_{i \rightarrow\infty} q_{i}.
\end{align}

Obviously, \eqref{eq:composed} depends on the edge-oriented degree distributions $\psi$, $\lambda$ and probabilities $\pi_j$.
While the latter depends on the physical layer operation, $\psi$ and $\lambda$ can be optimized through optimization of $\Psi$ and $\Lambda$, which in turn are optimized by optimizing 
$\beta$ and $\alpha$, see \eqref{eq:bard} and \eqref{eq:alpha}.
Specifically, in the proposed scheme, we optimize $\alpha$ and $\beta$ in order to maximize the expected throughput $\gamma$, see Section~\ref{sec:scheme_uplink}, which can be expressed as
\begin{align}
\gamma = \frac{p_d \, K}{ \Delta } R \frac{ L - \tau}{ \tau } = \frac{p_d}{\alpha} R ( L - \tau ).
\end{align}
\noindent We conclude by noting that the probabilities $\pi_j$, $j \geq 1$, are intractable to express analytically. Consequently, we evaluate $\pi_j$ using Monte Carlo simulations for the analytical results in the following section.



\section{Numerical Results} \label{sec:results}
This section presents numerical results based on simulations and evaluations of the analysis in section \ref{sec:analysis}. We compare the CPA scheme with the ALOHA random access scheme as described in connection with \eqref{eq:max} and a scheduled conventional massive MIMO scheme, referred to as SMM. The SMM scheme is assumed to guarantee interference free transmissions, i.e. $\left\vert{\mathcal{A}_n^j}\right\vert = 1$ $\forall$ $n,j$, thus not needing SIC. Users are assigned resources in a round-robin fashion. This scheme is considered an upper bound for a random access scheme.

The relevant parameters can be divided into two groups; system parameters and scheme parameters. System parameters are assumed to be given, whereas scheme parameters can be optimized for maximum throughput. We denote optimized parameters with a superscript $\star$, e.g. $\alpha^\star$ is the overhead factor, which maximizes the throughput. The throughput resulting from optimized parameters is denoted as $\gamma^\star$. All evaluations consider the case of $K=1000$, QPSK modulation and $\sigma_n^2=0.1$, i.e. an SNR of $10$~dB.

\begin{center}
\begin{tabular}{ |c|l|c|l| } \hline
  \multicolumn{2}{|c|}{\textbf{System Parameters}} & \multicolumn{2}{c|}{\textbf{Scheme Parameters}} \\ \hline \hline
  $\sigma_n^2$ & Noise power & $\tau$ & Pilot sequence length \\ \hline
  $K$ & Users in cell & $\alpha$ & Overhead factor \\ \hline
  $L$ & Coherence time & $\beta$ & Avg. factor node degree \\ \hline
  $M$ & Antennas at BS & $R$ & Channel code rate \\ \hline
\end{tabular}
\end{center}

\begin{figure}[t]
 \centering
 \includegraphics[width=1\columnwidth]{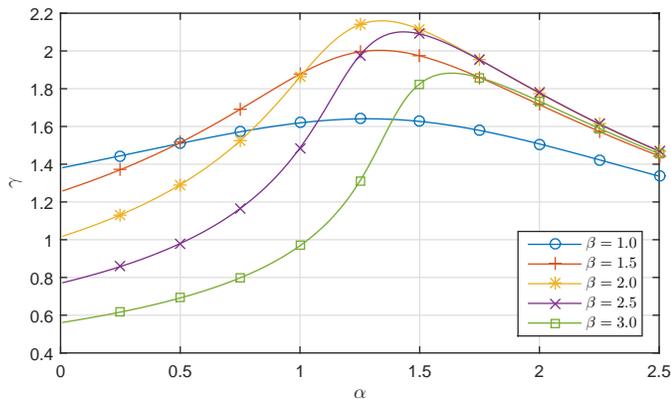}
 \caption{And-or tree evaluation of throughput as a function of $\alpha$ for different values of $\beta$. Fixed parameters are $\tau=4$, $L=64$, $M=400$ and $R=1$.}
 \label{fig:gamma_alpha}
\end{figure}

Initially, we present results on numerical optimizations of the parameters $\alpha$ and $\beta$. Fig.~\ref{fig:gamma_alpha} shows the throughput, computed using and-or tree evaluation, as a function of $\alpha$ for different values of $\beta$, with $\tau=4$, $L=64$, $M=400$ and $R=1$, i.e. no physical layer channel code. It is clear that the value of $\alpha$ has a great impact on the performance of the CPA scheme. Performance peaks at $\alpha$ slightly above 1, after which throughput decreases. This is the point at which most user messages can be resolved with SIC, relative to the invested overhead $\alpha$. Increasing $\alpha$ further, thus adding more resources to the frame, will just be a waste. 

\begin{figure}[t]
 \centering
 \includegraphics[width=1\columnwidth]{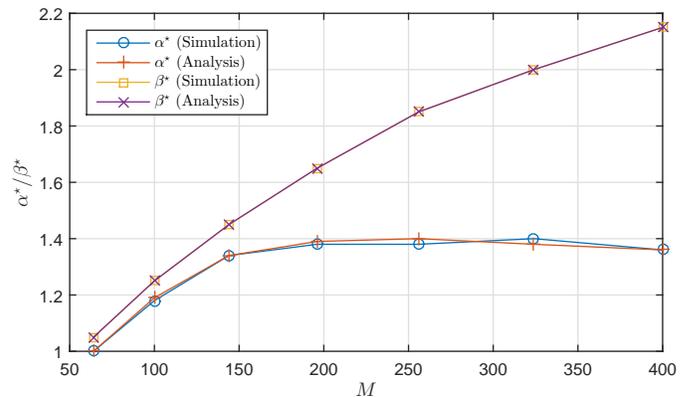}
 \caption{Optimal values of $\alpha$ and $\beta$ as a function of the number of antennas at the BS. Fixed parameters are $\tau=4$, $L=64$ and $R=1$.}
 \label{fig:ab_M}
\end{figure}

Fig.~\ref{fig:ab_M} shows the optimal values of $\alpha$ and $\beta$ as a function of $M$, with $\tau=4$, $L=64$ and $R=1$. Both analytical and simulation results are included and shown to correspond very well. It is seen that increasing $M$ allows for an increasing $\beta$, which indicates that SIC is better able to operate reliably. This is a result of improved orthogonality between user channels and improved temporal stability of the channel powers, which are essential properties as described in connection with equations \eqref{eq:fnj2} and \eqref{eq:gnj}.

\begin{figure}[t]
 \centering
 \includegraphics[width=1\columnwidth]{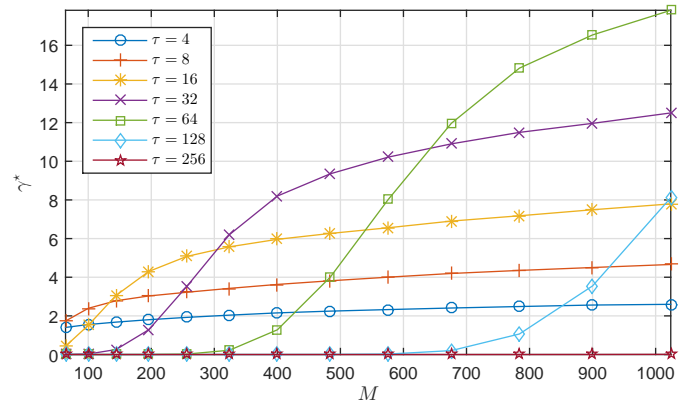}
 \caption{Throughput of the proposed CPA scheme as a function of the number of antennas at the BS for different values of $\tau$. Fixed parameters are $L=512$, $R=1$ and both $\alpha$ and $\beta$ have been numerically optimized for each data point.}
 \label{fig:gamma_M_tau_10}
\end{figure}

\begin{figure}[t]
 \centering
 \includegraphics[width=1\columnwidth]{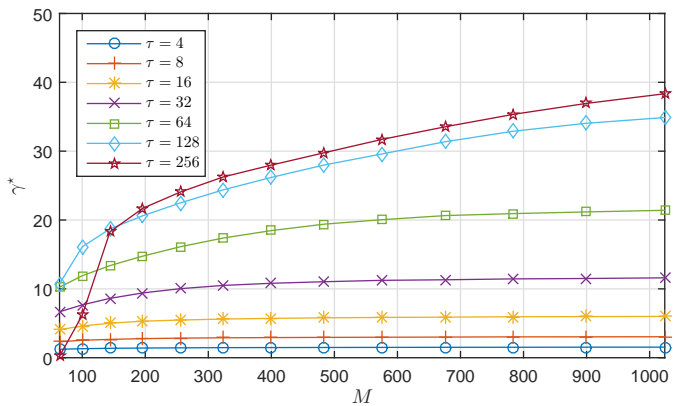}
 \caption{Throughput of the proposed CPA scheme as a function of the number of antennas at the BS for different values of $\tau$. Fixed parameters are $L=512$, $R=0.5$ and both $\alpha$ and $\beta$ have been numerically optimized for each data point.}
 \label{fig:gamma_M_tau_05}
\end{figure}

Next, we turn our attention to the choice of a well performing value of $\tau$. Note that $\alpha$ is proportional to the product of $\tau$ and $\Delta$, see \eqref{gamman}. Therefore, we now consider the optimal way to reach the desired $\alpha$ through the choice of $\tau$ and thereby $\Delta$. One the one hand, increasing $\tau$, while keeping $\alpha$ fixed, provides the same orthogonal resources using less time slots, which increases the potential throughput. On the other hand, increasing $\tau$ entails a rate loss, due to pilot symbols. Moreover, it entails a decrease in SINR, due to the increased number of orthogonal resources per time slot, which, for a given value of $\beta$, means an increased number of active users. This trade-off creates a correlation between the optimal $\tau$ and the value of $M$, as is seen in Fig.~\ref{fig:gamma_M_tau_10}, where results for $L=512$ and $R=1$ are plotted. Increasing $M$ compensates for the decrease in SINR, such that SIC can operate at a higher $\tau$. The throughput gain from this is seen to be quite significant. Fig.~\ref{fig:gamma_M_tau_05} illustrates the case where a rate $0.5$ channel code is applied at the physical layer. In this case, the scheme can cope with lower SINR levels, and thus higher values of $\tau$. Obviously, the drawback is the rate loss from the channel code. However, the gain of SIC overcompensates the rate loss from the channel code, whereby significantly higher throughput is achieved compared to uncoded operation.

\begin{figure}[t]
 \centering
 \includegraphics[width=1\columnwidth]{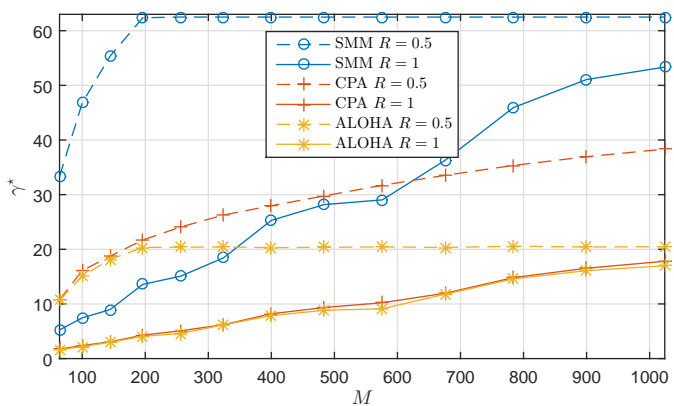}
 \caption{Comparison of the evaluated schemes at $R=0.5$ and $R=1$ and optimized values of $\tau$, $\alpha$ and $\beta$.}
 \label{fig:gamma_M_schemes}
\end{figure}

Finally, we present a comparison between the proposed CPA scheme and the two references, SMM and ALOHA. For all schemes, $\tau$, $\alpha$ and $\beta$ have been optimized. Fig.~\ref{fig:gamma_M_schemes} shows results for both $R=0.5$ and $R=1$ with $L=512$. As expected, the performance of all schemes increases with $M$. However, in the case of $R=0.5$, the ALOHA scheme experiences a saturation of the performance at roughly $M=200$, whereas the CPA scheme continues to increase. The reason is that the ALOHA scheme can only benefit from the increased SINR until the point, where degree one signals are decoded with high probability. The CPA is able to further benefit, due to improved SIC. Roughly a doubling of the throughput is achieved at $M=1024$ and $R=0.5$ compared to ALOHA, which closes a significant part of the gap to the upper bound given by scheduled operation.

\section{Conclusions}\label{sec:conclusions}
Crowd scenarios present a particularly challenging access problem in massive MIMO systems. The intermittent traffic from users and the scarcity of pilot sequences makes orthogonal scheduling infeasible. We presented a solution based on coded random access, which leverages on the channel hardening properties of massive MIMO. These allow us to view a set of contaminated pilot signals as a graph code on which iterative belief propagation can be performed. Using the tool of the and-or tree evaluation, we were able to analytically optimize the degree distribution of the random access code. With optimized parameters, the proposed solution proves highly efficient, comfortably outperforming the conventional ALOHA approach to random access.

\bibliographystyle{ieeetr}
\bibliography{bibliography}

\begin{thebibliography}{10}

\bibitem{Petar5G}
F.~Boccardi, R.~Heath, A.~Lozano, T.~Marzetta, and P.~Popovski, ``Five
  disruptive technology directions for 5g,'' {\em Communications Magazine,
  IEEE}, vol.~52, pp.~74--80, February 2014.

\bibitem{marzetta06}
T.~Marzetta, ``How much training is required for multiuser {MIMO}?,'' in {\em
  Signals, Systems and Computers, 2006. ACSSC '06. Fortieth Asilomar Conference
  on}, pp.~359--363, Oct 2006.

\bibitem{rusek13}
F.~Rusek, D.~Persson, B.~K. Lau, E.~Larsson, T.~Marzetta, O.~Edfors, and
  F.~Tufvesson, ``Scaling up {MIMO}: Opportunities and challenges with very
  large arrays,'' {\em Signal Processing Magazine, IEEE}, vol.~30, pp.~40--60,
  Jan 2013.

\bibitem{gesbert13}
H.~Yin, D.~Gesbert, M.~Filippou, and Y.~Liu, ``A coordinated approach to
  channel estimation in large-scale multiple-antenna systems,'' {\em Selected
  Areas in Communications, IEEE Journal on}, vol.~31, pp.~264--273, February
  2013.

\bibitem{ashikhmin12}
A.~Ashikhmin and T.~Marzetta, ``Pilot contamination precoding in multi-cell
  large scale antenna systems,'' in {\em Information Theory Proceedings (ISIT),
  2012 IEEE International Symposium on}, pp.~1137--1141, July 2012.

\bibitem{jose11}
J.~Jose, A.~Ashikhmin, T.~Marzetta, and S.~Vishwanath, ``Pilot contamination
  and precoding in multi-cell {TDD} systems,'' {\em Wireless Communications,
  IEEE Transactions on}, vol.~10, pp.~2640--2651, August 2011.

\bibitem{METIS}
A.~Osseiran, F.~Boccardi, V.~Braun, K.~Kusume, P.~Marsch, M.~Maternia,
  O.~Queseth, M.~Schellmann, H.~Schotten, H.~Taoka, H.~Tullberg, M.~Uusitalo,
  B.~Timus, and M.~Fallgren, ``Scenarios for 5g mobile and wireless
  communications: the vision of the metis project,'' {\em Communications
  Magazine, IEEE}, vol.~52, pp.~26--35, May 2014.

\bibitem{Icassp}
E.~de~Carvalho, E.~Björnson, E.~G. Larsson, and P.~Popovski, ``Random access
  for massive mimo systems with intra-cell pilot contamination.,'' {\em CoRR},
  vol.~abs/1509.08305, 2015.

\bibitem{SUCR}
E.~Bj{\"{o}}rnson, E.~de~Carvalho, E.~G. Larsson, and P.~Popovski, ``Random
  access protocol for massive {MIMO:} strongest-user collision resolution
  {(SUCR)},'' {\em CoRR}, vol.~abs/1512.00490, 2015.

\bibitem{Liva}
G.~Liva, ``{G}raph-{B}ased {A}nalysis and {O}ptimization of {C}ontention
  {R}esolution {D}iversity {S}lotted {ALOHA},'' {\em Communications, IEEE
  Transactions on}, vol.~59, pp.~477 --487, february 2011.

\bibitem{SPV2012}
C.~Stefanovic, P.~Popovski, and D.~Vukobratovic, ``{F}rameless {ALOHA} protocol
  for {W}ireless {Networks},'' {\em IEEE Comm. Letters}, vol.~16,
  pp.~2087--2090, Dec. 2012.

\bibitem{gcws}
J.~H. S{\o}rensen, E.~de~Carvalho, and P.~Popovski, ``Massive mimo for crowd
  scenarios: A solution based on random access,'' in {\em 2014 IEEE Globecom
  Workshops (GC Wkshps)}, pp.~352--357, Dec 2014.

\bibitem{PSLP2014}
E.~Paolini, C.~Stefanovic, G.~Liva, and P.~Popovski, ``{C}oded {R}andom
  {A}ccess: {H}ow {C}oding {T}heory {H}elps to {B}uild {R}andom {A}ccess
  {P}rotocols,'' {\em IEEE Commun. Mag.}, vol.~53, pp.~144--150, June 2015.

\bibitem{lt}
M.~Luby, ``{LT Codes},'' in {\em Foundations of Computer Science, 2002.
  Proceedings. The 43rd Annual IEEE Symposium on}, pp.~271--280, 2002.

\bibitem{sorensen}
J.~S{\o}rensen, P.~Popovski, and J.~{\O}stergaard, ``Design and analysis of lt
  codes with decreasing ripple size,'' {\em Communications, IEEE Transactions
  on}, vol.~60, pp.~3191 --3197, november 2012.

\bibitem{LMS1998}
M.~G. Luby, M.~Mitzenmacher, and A.~Shokrollahi, ``{A}nalysis of {R}andom
  {P}rocesses via {A}nd-{O}r {T}ree {E}valuation,'' in {\em Proc. of 9th
  ACM-SIAM SODA}, (San Francisco, CA, USA), Jan. 1998.

\bibitem{MCT}
T.~Richardson and R.~Urbanke, {\em {M}odern {C}oding {T}heory}.
\newblock Cambridge University Press, Cambridge, UK, 2007.

\bibitem{SMP2014}
C.~Stefanovic, M.~Momoda, and P.~Popovski, ``{E}xploiting {C}apture {E}ffect in
  {F}rameless {ALOHA} for {M}assive {W}ireless {R}andom {A}ccess,'' in {\em
  IEEE WCNC 2014}, (Istanbul, Turkey), Apr. 2014.

\end{thebibliography}
\end{document}